# Double-slit time diffraction at optical frequencies

Romain Tirole[1]*, Stefano Vezzoli[1]*, Emanuele Galiffi[2], Iain Robertson[1], Dries Maurice[1], Benjamin Tilmann[3], Stefan A. Maier[1,3,4], John B. Pendry[1], Riccardo Sapienza[1]

[1] The Blackett Laboratory, Department of Physics, Imperial College London; London SW7 2BW, United Kingdom

[2] Photonics Initiative, Advanced Science Research Center, City University of New York; 85 St. Nicholas Terrace, 10031, New York, NY, USA

[3] Chair in Hybrid Nanosystems, Nanoinstitut München, Ludwig-Maximilians-Universität München; 80539 München, Germany

[4] School of Physics and Astronomy, Monash University; Clayton Victoria 3800, Australia

*These authors contributed equally to this work.

**Abstract**

The wave nature of light is revealed by diffraction from physical structures. We report a time-domain version of the classic Young's double-slit experiment: a beam of light twice gated in time produces an interference in the frequency spectrum. The 'time slits', narrow enough to produce diffraction at optical frequencies, are generated from a thin film of Indium-Tin-Oxide illuminated with high-power infrared pulses, inducing a fast reflectivity rise, followed by a slower decay. Separation between the time slits determines the period of oscillations in the frequency spectrum, while the decay of fringe visibility in frequency reveals the shape of the time slits. Here we find a surprise: many more oscillations are visible than expected from existing theory, implying a rise time for the leading edge of around 1-10 fs, approaching an optical cycle of 4.4 fs. This is over an order of magnitude faster than the width of the pump and can be inferred from the decay of the frequency oscillations.

**Main**

In 1802 Thomas Young performed the renowned double-slit experiment – a wave transmitted through a spatial double thin aperture – to determine if light was made of particles or waves. The observed diffraction pattern provided the incontrovertible signature of the interference between emerging spherical waves from each spatial slit. Since the first experiment with light waves *(1)*, the double-slit experiment has revealed the wave-particle duality of quantum objects, such as single photons *(2)*, electrons *(3,4)*, neutrons *(6)*, atoms *(7)* and large molecules *(8)*. Yet, the temporal counterpart of Young's double-slit experiment – a wave interacting with a double temporal modulation of an interface *(8)* – remains elusive.

Wave-matter interaction in time-varying media exhibits strikingly different dynamics than in conventional passive media, as energy can be exchanged between wave and medium, the wave frequency changes, and the propagation is no longer symmetric under an inversion of time. Amidst the recent explosion of studies on optical time-varying media *(9)*, time refraction *(10-12)* and time reversal *(13)*, have been achieved in epsilon-near-zero (ENZ) semiconductors, as for example Indium-Tin-Oxide (ITO) *(14)(15)*, modulated by ultrafast laser pulses.

Here, we report the observation of the temporal analog of the double-slit experiment for light waves, showing a clear signature of spectral oscillations for the time-diffracted light, and an inversely proportional relation between slit separation and the period of the oscillations. Moreover, the observed oscillations serve as a sensitive probe of the ITO response time, which we measure to be of the order of an optical cycle, much faster than previously thought.

In the conventional Young's double-slit experiment, the diffracted pattern has a characteristic oscillatory profile, with minima corresponding to momenta **k** for which destructive interference suppresses wave propagation, as illustrated in Fig. 1A. These minima have a separation in *momentum* space that increases for decreasing *spatial* separation of the slits. In the Fraunhofer approximation of diffraction, i.e. for aperture size much smaller than the observation distance, the field distribution in the far field can be approximated as the Fourier transform $\bar{A}(k_x)$ of the aperture function A(x) (Fig. 1B). The far field interference pattern can be plotted in a dispersion diagram, where it is described as horizontal transitions (Fig. 1C), with new modes appearing, characterized by different momenta but the same frequency, as required by temporal translational invariance.

In the temporal domain, the corresponding scenario, a *temporal* double-slit, consists of a deeply subwavelength slab characterized by a time-varying dielectric function which is an aperture function A(t) defining two time slits (Fig. 1D). In this case the wave will be time-diffracted into a *frequency* spectrum of frequencies $\bar{A}(\omega)$, the Fourier transform of the aperture function A(t), around the incident carrier frequency (Fig. 1E), whereas the in-plane momentum $k_x$ will be conserved by translational symmetry (Fig. 1F).

Experimentally, we create time slits by inducing an ultrafast change in the complex reflection coefficient of a time-varying mirror *(16)* made of a 40 nm thin film of ITO, with epsilon-near-zero (ENZ) frequency of 227 THz (1320 nm). The film is deposited on a glass coverslip and covered by a 100 nm layer of gold to improve field confinement and reflectivity, as sketched in Fig. 1G. The time modulation is driven by two 225 fs (FWHM) pulses at 230.2 THz (1300 nm), impinging near the Berreman angle (60°), ensuring efficient field coupling and an enhanced nonlinear response in ITO (see Supplemental Information SI for more details). We measure the temporal double-slit by a pump-probe experiment (Fig. 1H blue line). The reflectivity of the sample changes from 0.08 to 0.6, giving a modulation contrast of 76% for pump intensities around 124 GW/cm$^2$, large enough to saturate the mirror response (see Fig. S3C). The measured reflectivity has the shape of a double time slit, with a fast rise time and a slower decay, given by a convolution of the driving pump pulses, the probe and the material response. The time slits separation can be tuned by adjusting the relative delay of the two pump pulses via a delay line.

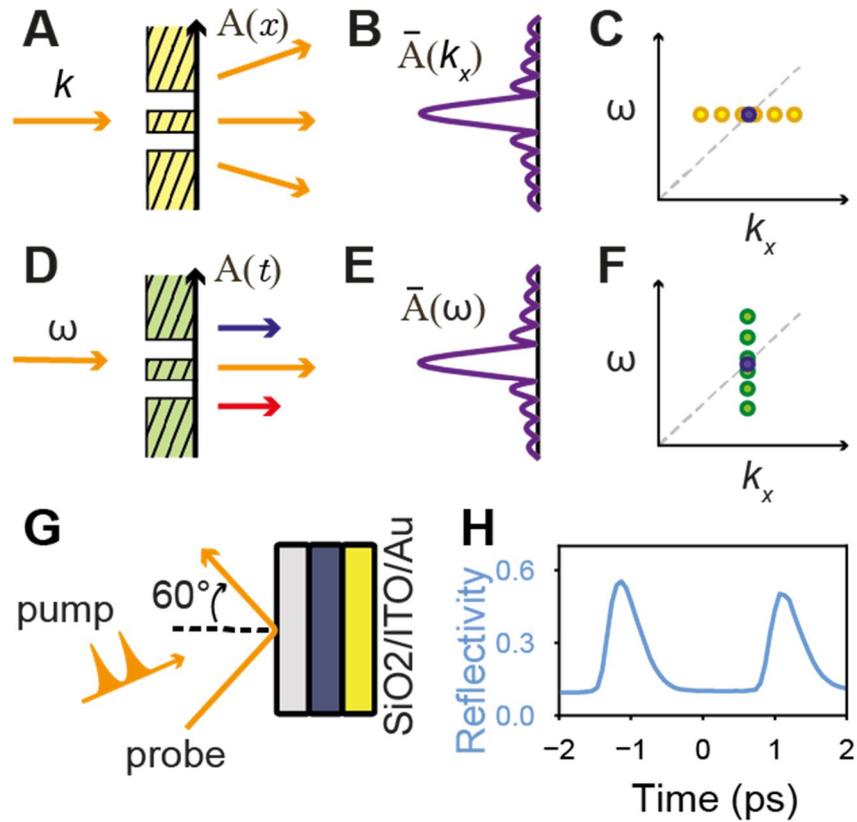

Figure 1. **Concept and realization of the double-slit diffraction experiment in time. (A)** Conventional spatial double-slit experiment: as light diffracts from a spatial double slit, **(B)** the aperture changes the beam's in-plane momentum $k_x$, corresponding to **(C)** a horizontal transition in the dispersion diagram. **(D)** Temporal double-slit experiment: as light interacts with a double time modulation, **(E)** an aperture in time acts on the frequency ω of the beam. **(F)** The transition is now vertical in the dispersion diagram. **(G)** Experimental realization**:** pump and probe beams are incident close to 60 deg onto a 40 nm ITO slab on glass, coated with a 100 nm gold film. **(H)** Temporal change of the sample reflectivity (blue line) with a 2.3 ps separation between the slits.

Direct evidence of time diffraction from the temporal double slit is given by illuminating the sample with a probe pulse (230.2 THz carrier frequency, 1 THz bandwidth) of duration 794 fs (FWHM) and by monitoring the reflected probe spectrum. The probe pulse is spectrally broadened, exhibiting new frequency content up to ~10 bandwidths away from the carrier frequency. A clear spectral modulation, with sinusoidal oscillations, is evident in Fig. 2A,B (red lines), shown for two different time slit separations. Measurements with 800 fs slit temporal separation (Fig. 2A) present much faster oscillations than with 500 fs (Fig. 2B). The separation of the time slits determines the period of oscillations in the frequency spectrum, while the shape of each time slit constrains the oscillation decay and therefore the number of oscillations that are visible. Up to six oscillations in the spectrum are evident in Fig. 2A,B, with an overall decay in intensity away from the central frequency of the probe.

These spectral oscillations are very well captured by a simple diffraction model (purple line in Fig. 2A,B). This model calculates the spectral evolution of the probe from the Fourier transform of the product of the time-varying reflection coefficient $r(t)$, i.e. the slit aperture function, and the probe pulse temporal profile. In such a model, the slit separation and the decay time of each slit are fixed by the pump/probe data (Fig. S3), whereas the rise time is taken as a fitting parameter.

The oscillation decay with increasing change in frequency depends on the shape of the time slits, and in particular their rise time. Therefore, the measured spectral oscillations act as a very precise measurement of the ITO response time, with a resolution well beyond that of a conventional pump-probe experiment. Here, we find unexpected physics: many more oscillations are visible than expected from existing theory, implying a rise time for the leading edge estimated to be 1-10 fs, i.e. of the order of an optical cycle (4.4 fs); this is much faster than previously thought *(10)-(12)*, although recent work has shown evidence of a speeding up of the response time for pump intensities beyond the linear regime *(16)*. Moreover, the oscillations are very close to the asymptotic limit of an ideal time slit with a Heaviside (infinitely fast) rise profile, for which the amplitude of the oscillations is expected to decay as $1/(f-f_0)^2$, where f is the frequency of oscillation and $f_0$ the probe central frequency (230.2 THz). This is evident when the spectrum is rescaled by the inverse frequency squared, as its intensity is almost constant (Fig. 2C-D). The dashed purple line in Fig. 2C-D is the asymptotic theory with two Heaviside time slits (see Fig. S2).

The signature of time diffraction is further revealed by the period of these oscillations which is inversely proportional to the slit separation, as plotted in Fig. 2E. The agreement between the measured data (open circles) and the model is remarkable. The error bars show the accuracy of the slit separation (50 fs) and period (0.15 THz). The shaded area identifies time slit separations smaller than 300 fs where the two time slits start to merge, and the oscillations are not visible anymore.

Further insight into the temporal diffraction process can be reached by analyzing the full interferogram of the time diffracted light as a function of the slit separation, as in Fig. 2F. Oscillations appear more pronounced on the red side of the spectrum with frequencies as far out as 10 THz (~ 60 nm) while only exhibiting frequencies 4 THz away on the blue side (on top of the probe pulse initial spectral content of 1 THz width). The asymmetric interferogram is explained by the time evolution of the phase of the complex reflection coefficient, causing a Doppler shift of the spectrum, often dubbed time refraction *(10)* (see Fig. S4). The theoretical plot in Fig. 2G does not capture this asymmetry, as it does not include a phase change during reflection.

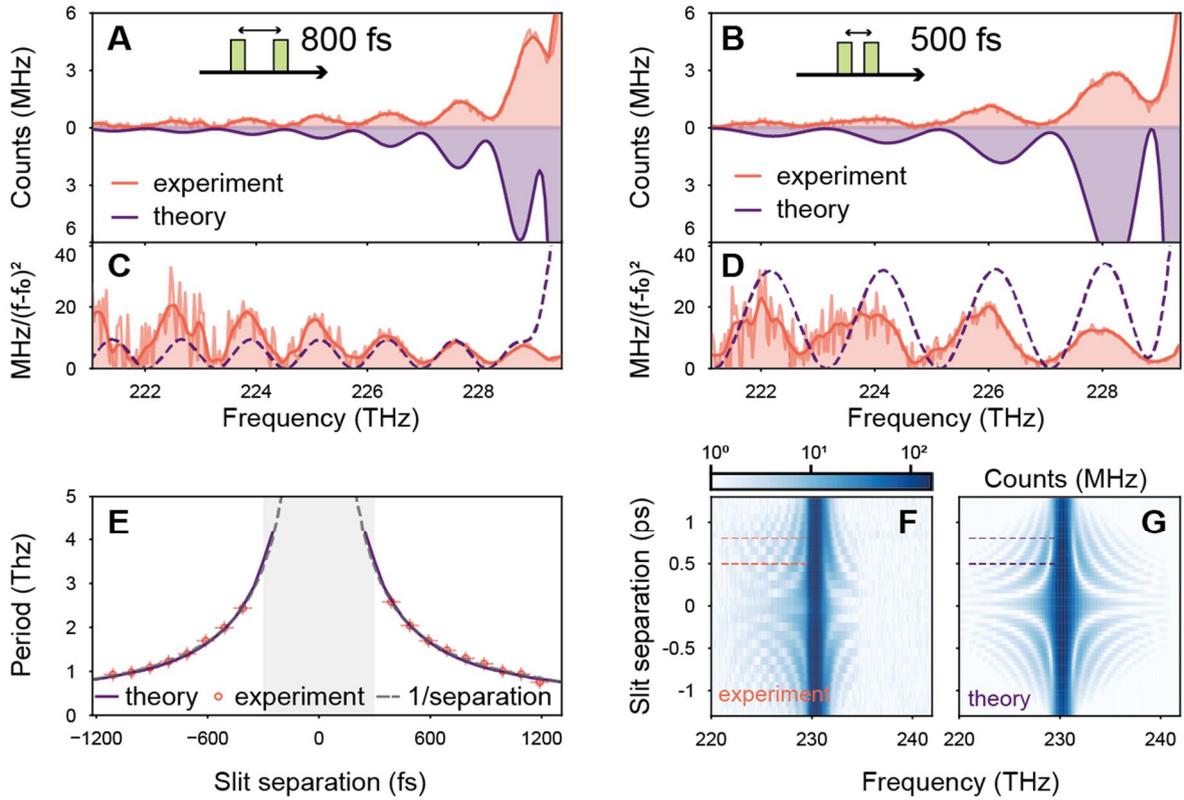

Figure 2. **Observation of a spectral diffraction pattern from temporal double-slits. (A,B)** Oscillations in the reflection spectrum of a 794 fs probe pulse interacting with the double-slit temporal aperture: experiment (light red curve is raw data, dark red curve is the smoothed data) against theoretical model (purple line). The oscillations are visible up to 10 THz away from the incident frequency (230.2 THz, 1 THz bandwidth). **(C,D)** Experimental oscillations rescaled by $(f-f_0)^2$, and asymptotic theoretical curve (dashed purple line). The measured oscillation period of the spectral diffraction is plotted in **(E)** (red circles) as a function of slit separation, in very good agreement with the period extracted from our model (purple line) and the inverse slit separation dependence (gray dashed line). The grey area highlights separations smaller than 300 fs. **(F,G)** Full interferograms of the time diffracted light as a function of slit separation and frequency, (F) experiment and (G) theory. The dashed lines show the range where data in (A-D) are shown.

Time diffraction of waves has many similarities with its spatial counterpart, such as a sinusoidal profile with an inverse dependence on the time slit separation, but also differences, related to the rise-decay asymmetry of the time slit and the Doppler shift of the diffracted light. Time diffraction requires a time slit modulation fast enough to develop frequency oscillations that decay slowly, over frequencies larger than the probe bandwidth, as we show here.

In conclusion, we report a direct observation of spectral oscillations, at optical frequencies, resulting from double-slit time diffraction, the temporal analogue of the Young's slits experiment. The measurements show a clear inverse proportionality between the oscillation period and the time slit separation. These oscillations reveal a 1-10 fs temporal dynamics of the ITO/Au bilayer, much faster than previously thought and beyond the adiabatic and linear intensity dependence assumed so far in most theoretical models *(10)(11),(17)*, calling for a new fundamental understanding of such ultrafast non-equilibrium responses. The observation of temporal Young's double-slit diffraction paves the way for optical realizations of time-varying metamaterials, promising enhanced wave functionalities such as nonreciprocity *(18)*, new forms of gain *(19)(20)*, time reversal *(21)(22)* and optical Floquet topology *(23)(24)*. Double-slit time diffraction could be extended to other wave domains, e.g. matter waves (6), optomechanics *(25)* and acoustics (26)(27), electronics *(28)*, and spintronics *(29)*, with applications for pulse shaping, signal processing and neuromorphic computation. Finally, the visibility of the oscillations can be used to measure the phase coherence of the wave interacting with it, similarly to matter-wave interferometers *(6)*.

**Data and materials availability**

Data and codes are available upon request.

**References**

(1) T. Young, *A Course of Lectures on Natural Philosophy and the Mechanical Arts*. (J. Johnson, London, 1807)
(2) G. I. Taylor, Interference fringes with feeble light. *Proc. Camb. Philos. Soc.* **15** 114-115 (1909) C.
(3) Jönsson, Elektroneninterferenzen an mehreren künstlich hergestellten Feinspalten. *Zeitschrift für Physik* **161** 454-474 (1961)
(4) P. G. Merli, G. F. Missiroli, G. Pozzi, On the statistical aspect of electron interference phenomena. *American Journal of Physics* **44** 306-7 (1976)
(5) A. Zeilinger, R. Gähler, C.G.Shull, W. Treimer, W Mampe, Single and double-slit diffraction of neutrons. *Rev. Mod. Phys*. **60**, 1067-1073 (1988)
(6) D. Cronin, J. Schmiedmayer, D. E. Pritchard, Optics and interferometry with atoms and molecules. *Rev. Mod. Phys.* **81** 1051–1129 (2009)
(7) K. Hornberger, S. Gerlich, P. Haslinger, S. Nimmrichter, M. Arndt, Colloquium: Quantum interference of clusters and molecules. *Rev. Mod. Phys.* **84** 157–173 (2012)
(8) M. Moshinsky. Diffraction in time. *Phys. Rev.* **88** 625–631 (1952)
(9) E. Galiffi, R. Tirole, S. Yin, H. Li, S. Vezzoli, P. A. Huidobro, M. G. Silveirinha, R. Sapienza, A. Alù, J. B. Pendry, Photonics of time-varying media. Adv. Photonics 4 01 (2022)
(10) Y. Zhou, M. Z. Alam, M. Karimi, J. Upham, O. Reshef, C. Liu, A. E. Willner, R. W. Boyd, Broadband frequency translation through time refraction in an epsilon-near-zero material. *Nat. Commun.*, **11** 1 (2020).
(11) V. Bruno, S. Vezzoli, C. DeVault, E. Carnemolla, M. Ferrera, A. Boltasseva, V. M. Shalaev, D. Faccio, M. Clerici, Broad frequency shift of parametric processes in epsilon-near-zero time-varying media. *Appl. Sci.* **10** 4 1318 (2020)
(12) J. Bohn, T. S. Luk, S. Horsley, E. Hendry. Spatiotemporal refraction of light in an epsilon-near-zero indium tin oxide layer: frequency shifting effects arising from interfaces. *Optica* **8** 12 1532 (2021)
(13) S. Vezzoli, V. Bruno, C. DeVault, T. Roger, V. M. Shalaev, A. Boltasseva, M. Ferrera, M. Clerici, A. Dubietis, D. Faccio, Optical time reversal from time-dependent epsilon-near-zero media. *Phys. Rev. Lett.* **120** 4 043902 (2018)
(14) N. Kinsey, C. DeVault, A. Boltasseva, V. M. Shalaev, Near-zero-index materials for photonics. *Nat. Rev. Mater.* **4** 12 742–760 (2019)

(15) M. Z. Alam, I. De Leon, R. W. Boyd, Large optical nonlinearity of indium tin oxide in its epsilon-near-zero region. *Science* **352** 6287 795–797 (2016)

(16) R. Tirole, E. Galiffi, K. Dranczewski, T. Attavar, B. Tilmann, Y.-T. Wang, P. A. Huidobro, A. Alù, J. B. Pendry, S. A. Maier, S. Vezzoli, R. Sapienza, Saturable time-varying mirror based on an ENZ material. *ArXiv e-prints* (2022)

(17) J. Bohn, T. S. Luk, C. Tollerton, S. W. Hutchings, I. Brener, S. Horsley, W. L. Barnes, E. Hendry, All-optical switching of an epsilon-near-zero plasmon resonance in indium tin oxide. *Nat. Commun.* **12** 1 (2021)

(18) D. L. Sounas, A. Alù, Non-reciprocal photonics based on time modulation. *Nat. Photonics*, **11** 12 774–783 (2017)

(19) T. T. Koutserimpas, R. Fleury, Nonreciprocal gain in non-hermitian time-floquet systems. *Phys. Rev. Lett.* **120** 8 087401 (2018)

(20) E. Galiffi, P. Huidobro, J. B. Pendry, Broadband nonreciprocal amplification in luminal metamaterials. *Phys. Rev. Lett.* **123** 20 206101 (2019)

(21) G. Lerosey, J. de Rosny, A. Tourin, A. Derode, G. Montaldo, M. Fink, Time reversal of electromagnetic waves. *Phys. Rev. Lett.* **92** 193904 (2004)

(22) V. Bruno, C. DeVault, S. Vezzoli, Z. Kudyshev, T. Huq, S. Mignuzzi, A. Jacassi, S. Saha, Y. D. Shah, S. A. Maier, D. R. S. Cumming, A. Boltasseva, M. Ferrera, M. Clerici, D. Faccio, R. Sapienza, V. M. Shalaev, Negative refraction in time-varying strongly coupled plasmonic-antenna–epsilon-near-zero systems. *Phys. Rev. Lett.* **124** (4) 043902 (2020)

(23) E. Lustig, Y. Sharabi, M. Segev, Topological aspects of photonic time crystals. *Optica* **5** 11 1390 (2018)

(24) A. Dutt, Q. Lin, L. Yuan, M. Minkov, M. Xiao, S. Fan, A single photonic cavity with two independent physical synthetic dimensions. *Science* **367** 6473 59–64 (2020)

(25) J. del Pino, J. J. Slim, E. Verhagen, Non-hermitian chiral phononics through optomechanically induced squeezing. *Nature* **606** 7912 82–87 (2022)

(26) R. Fleury, D. L. Sounas, C. F. Sieck, M. R. Haberman, A. Alù, Sound isolation and giant linear nonreciprocity in a compact acoustic circulator. *Science* **343** 6170 516–519 (2014)

(27) C. Cho, X. Wen, N. Park, J. Li, Digitally virtualized atoms for acoustic metamaterials. *Nat. Comm.* **11** 1 (2020)

(28) A. Nagulu, N. Reiskarimian, H. Krishnaswamy, Non-reciprocal electronics based on temporal modulation. *Nat. Electron.* **3** 5 241–250 (2020)

(29) K. Schultheiss, N. Sato, P. Matthies, L. Körber, K. Wagner, T. Hula, O. Gladii, J. E. Pearson, A. Hoffmann, M. Helm, J. Fassbender, H. Schultheiss, Time refraction of spin waves. *Phys. Rev. Lett.* **126** 13 137201 (2021)

**Supplementary Materials**

**Temporal and spectral characterization of the probe and pump pulses**

All pulses are generated by a Light Conversion Pharos solid state laser, set to 230.2 THz with a Light Conversion Orpheus optical parametric amplifier, with a nominal full width at half maximum (FWHM) of the electric field envelope of 225 fs. The pulses are roughly Gaussian in shape.

For the double slit experiment a probe pulse identical to the pump pulses is sent through a 4-f system, composed of a diffraction grating, a variable aperture and two lenses, which separates, filters and recombines different spectral components to broaden the pulse in time, in order to make it longer than the time slits separation. By selecting the spectral content of the probe beam with an aperture (pink line in Fig. S1A, original spectrum in beige for reference), the pulse is significantly extended in time. This is quantified by using cross-correlation measurements, where we collect the sum-frequency-generated (SFG) signal originating from the spatial overlap of the probe and the pump pulses in a 400 μm thick Gallium Phosphide crystal with second order nonlinearity.

The measured SFG signal corresponding to the pink line spectrum shown in Supp.Fig.1A is given in Fig. S1B (blue line). As the analytical solution to the cross-correlation measurement is known for a Gaussian pulse, we fit the FWHM of the probe pulse amplitude assuming such a shape (Supp. Fig. 1B red line). The agreement between theory and experiment shows that this is a well-suited approximation for the main peak of the probe envelope. For reference, the pristine pump pulse autocorrelation measurement and fit are also shown (blue and red dashed lines). The probe pulse is measured to have a 794 ± 37 fs FWHM, while the original pump pulse is recorded to have a 225 ± 5 fs FWHM, close to device specifications. These values refer to the electric field amplitude of the pulse, the corresponding intensity FWHM are $\sqrt{2}$ shorter.

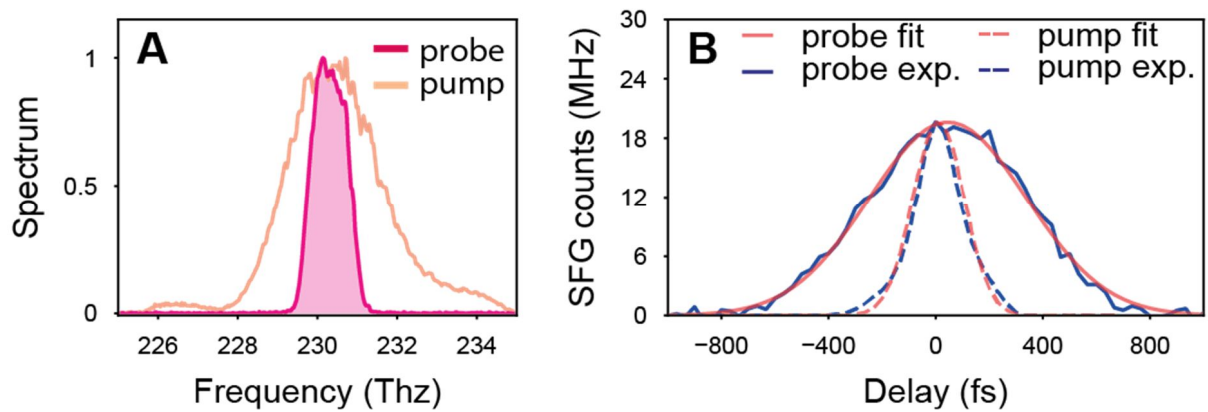

**Supplementary Figure 1**. Characterization of the probe pulse temporal width. **(A)** Measured original pump and probe spectrum (beige) and filtered probe spectrum (pink) used for the double slit experiment. **(B)** Cross-correlation of the probe-pump pulses obtained from sum-frequency generation (SFG) as a function of probe to pump delay: experimental (continuous blue) against fitted analytical theory for a gaussian pulse (continuous red). The autocorrelation of the pristine pump is shown for comparison (blue dashed for experiment, red dashed for theory), normalized to the intensity of the probe cross correlation signal.

**Modelling of the slits**

We model a single slit aperture function as:

$$f_{ss}(t) = \frac{1}{(1+e^{-\alpha t}) \times (1+e^{\beta t})}$$

where $\alpha$ and $\beta$ are taken to be positive constants. The double slit modulation profile of the amplitude reflection coefficient can then be expressed as

$$r(t) = A \times f_{ss}(t - \frac{S}{2}) + B \times f_{ss}(t + \frac{S}{2}) + C$$

where $S$ is the slit separation, $A$ and $B$ are the respective slit amplitudes and $C$ is a constant corresponding to the unmodulated reflection coefficient of the time-varying mirror. This function is normalized by fitting it to the measured intensity reflectivity change $R(t) = |r(t)|^2$ (Supp. Fig. 2A and Fig. S3A).

In Fig. 2C,D of the manuscript we compare the decay of the frequency oscillations to an asymptotic case, where the rise time becomes infinitely fast and the decay infinitely slow (that is when $\alpha \to +\infty$ and $\beta \to 0$). We model this limiting case using two consecutive Heaviside functions (denoted $H(t)$), as shown in Supp. Fig. 2B, using the following expression:

$$r(t) = A \times H(t - \frac{S}{2}) + B \times H(t + \frac{S}{2}) + C$$

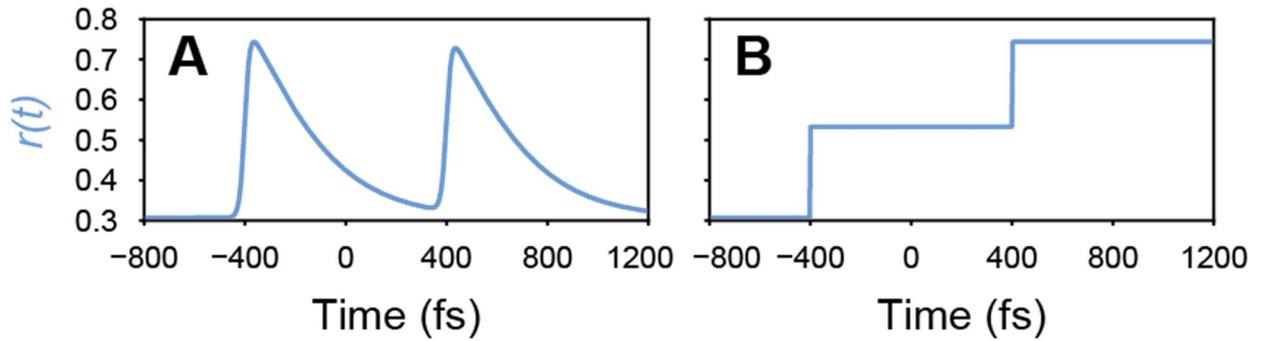

**Supplementary Figure 2**. Modelled reflection coefficient **(A)** Amplitude reflection coefficient $r(t)$ corresponding to a realistic modulation of the time-varying mirror (see Fig. S3A). **(B)** Amplitude reflection coefficient in the asymptotic limit.

**Characterization of the temporal double slit**

In order to characterize the double slit time modulation, we use a short 225 fs FWHM probe in a degenerate pump-probe experiment. The probe is incident on the sample at a 60° angle to the normal, while the pump beams are incident at a 6° angle difference on either side of the probe. All beams are p-polarized to couple them to the Berreman resonance. The probe and one of the pumps go through a delay line to control the probe arrival time on the sample and the temporal separation between the two. The reflected probe signal is sent to a Princeton Instruments NIRvana camera for spectral characterization. This is the same configuration of the double slit experiment illustrated in Fig. 2, except there the probe pulse is temporally broadened by passing it through the 4-f system.
We refer to our previous work *(16)* for a complete characterization of the ITO/Au sample's linear and nonlinear properties.

The double slit aperture function $R(t)$ can be measured by pump probe spectroscopy, as shown in Fig. S3A (blue line), where the probe reflectivity is plotted as a function of the delay for a fixed slit separation. The theoretical fit (dashed red line) comes from the amplitude reflection coefficient $r(t)$ defined in the previous section, convoluted with the envelope of the probe pulse in time $E_{probe}(t)$ (approximated to a Gaussian profile with 225 fs FWHM). The predicted change in intensity reflectivity is then $r(t)^2 * E_{probe}(t)^2$.

In addition to fixing the values of normalization constants $A$, $B$ and $C$, the fitting of the pump/probe reflectivity measurements allows us to assign a value for the $\beta$ coefficient at 1/400 fs$^{-1}$, corresponding to a decay time of the intensity reflectivity (defined as the time to decrease from 90% to 10%) of 625 fs (or 1/e of 330 fs, compatible with the literature *(17)*). From this measurement we can also set the relative amplitude of the second peak of the time slit to be 0.93% of the first one. As the rise time is limited by the probe pulse, we cannot estimate a precise value for it from this measurement and instead spectral data as in Fig. 2 are necessary.

Our model reproduces well the rise and decay time of the modulation but does not account for the short plateau at the maximum of the modulation, which arises when pumping near or beyond the saturation of the mirror's temporal response *(16)*. As time-varying effects depend exclusively on the slope of the modulation, we focus on reproducing this behavior rather than the extent of the flat region at the maximum.

A wide range of separations can be achieved by varying the delay between the two pumps (horizontal axis in Fig. S3B) while the delay of the probe is varied to map the modulation (vertical axis in Fig. S3B).

An intensity dependence study of the modulation is illustrated in Fig. S3C, showing that the double slit experiments are performed where the modulation contrast of the time-varying mirror is saturated with pump power. The two pumps have similar effect on the reflectivity change of the sample, although they illuminate it from different angles, 54° and 66° respectively, since the Berreman resonance is angularly wide *(16)*.

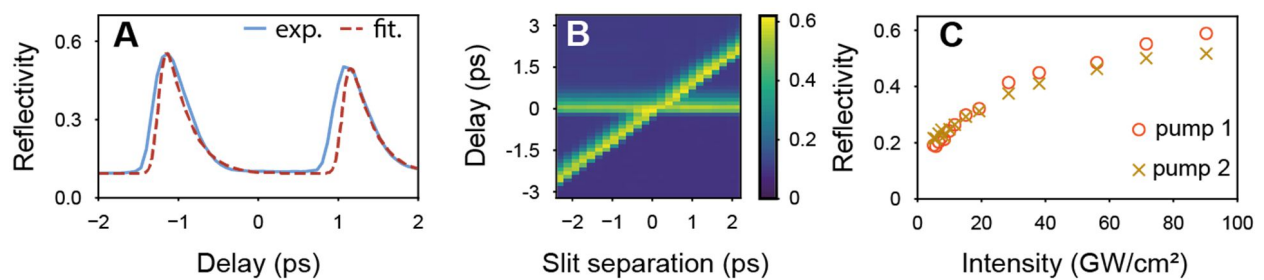

**Supplementary Figure 3**. Characterization of the temporal double slit with a short (225 fs) probe pulse. **(A)** Experimental intensity reflectivity (blue line) for a 2.3 ps separation between the time slits, as a function of the probe delay. This is fitted with the model in Fig. S2A (dashed red line). **(B)** Reflectivity for a linear variation of the slit separation (horizontal axis) measured as a function of the probe delay with one of the pumps (vertical axis). **(C)** Intensity dependence of the maximum achievable reflectivity for the two time slits driven by pump 1 (incident at 54°) and pump 2 (incident at 66°). The reflectivity saturates reaching a maximum value of 0.6.

**Time diffraction model**

The probe spectra in the double slit experiment are modelled as Fourier transforms of the product of the double slit reflection coefficient $r(t)$ and the probe field amplitude $E_{probe}(t)$ after the 4-f system (product of a Gaussian envelope with 794 fs FWHM and a carrier frequency 230.2 THz). The resulting spectrum $I(f)$ is then squared to return the measured field intensity:

$$I(f) = \mathscr{F}^2\left(r(t) \cdot E_{probe}(t - t_0)\right)$$

where $t_0$ is a parameter to account for a possible offset of the probe pulse envelope peak and the temporal center of the two slits. This model neglects the dispersion of the complex reflection coefficient of the sample, as it assumes the same temporal response at all frequencies. It also neglects the time evolution of the phase of $r(t)$ which is in principle a complex function, but is assumed here real.

Although this model reproduces well the period and amplitude of the oscillations of the double slit experiment, it fails to capture the asymmetry observed in the experimental spectrum, where more oscillations are visible on the red side (comparing Fig. S4A and B). We attribute this to the time evolution of the phase of the complex reflection coefficient, which causes a Doppler shift of the spectrum, often dubbed time refraction *(10)*. A phenomenological model of the phase evolution is known, as the pump/probe data are only sensitive to the amplitude of the reflectivity.
In order to understand better this asymmetry, we also model time diffraction using an adiabatic time-varying model which accounts for the material dispersion as well (see Fig. S4C). The permittivity modulation is computed by calculating a negative shift in the plasma frequency from the convolution of the gaussian pump intensity with an exponential rise and decay response function *(16)(17)*. This model of the pump-induced modulation in ENZ materials can reproduce the spectral oscillations, period and the interferogram asymmetry observed in Fig. S4A, however it fails to capture the full spectral extent of the oscillations because of the fast rise time of the experimental slit aperture function, which in our adiabatic simulations is fundamentally limited by the pump duration *(10)(12)*. This discrepancy highlights the need for new models to describe such non-adiabatic modulation.

The diffraction model can be used to fit the experimental spectral data, giving a value for the parameter $\alpha$ of 1/2 fs$^{-1}$, corresponding to an intensity reflectivity rise time (10-90%) of 7 fs. As shown in Fig. S4D, the data are close to the asymptotic limit of an infinitely fast rising slit, therefore the fitting is equally good for a reflectivity rise time (10-90%) in the range 1-10 fs.

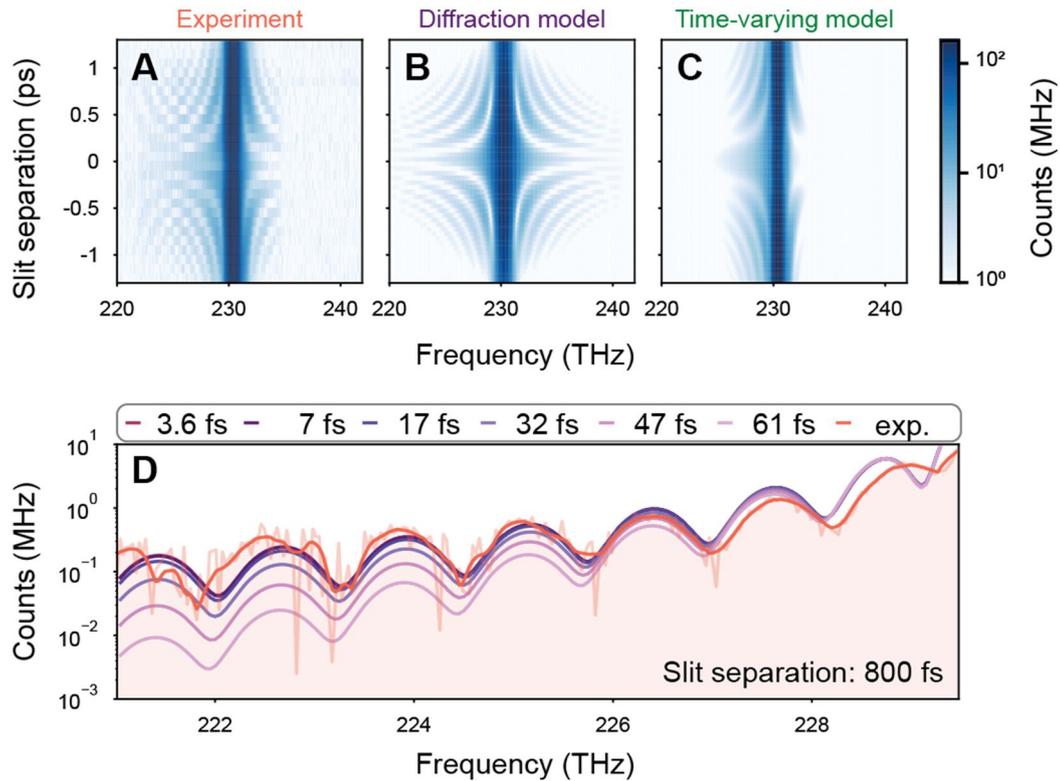

**Supplementary Figure 4**. Comparison of the models against experimental data. (**A-C**) Signal against frequency and slit separation for (**A**) experiment, (**B**) the time diffraction model and (**C**) the adiabatic, dispersive time-varying model. The plots are color-saturated to ensure a fair quantitative comparison between the fringe visibility of the respective datasets. (**D**) Experimental oscillation spectrum on a logarithmic scale at a slit separation of 800 fs compared to the theoretical one for various material response times.